# Direct observation of enhanced electron-phonon coupling in copper nanoparticles in the warm-dense matter regime

Quynh L. D. Nguyen[1]*[†], Jacopo Simoni[2,7], Kevin M. Dorney[1], Xun Shi[1], Jennifer L. Ellis[1], Nathan J. Brooks[1], Daniel D. Hickstein[3], Amanda G. Grennell[4], Sadegh Yazdi[5], Eleanor E. B. Campbell[6], Liang Z. Tan[7], David Prendergast[7], Jerome Daligault[2], Henry C. Kapteyn[1,3], Margaret M. Murnane[1]

[1]*JILA, Department of Physics, University of Colorado and NIST, Boulder, Colorado 80309, USA*
[2]*Theoretical Division, Los Alamos National Laboratory, NM 87545, USA*
[3]*KMLabs. Inc., 4775 Walnut St #102, Boulder, CO 80301, USA*
[4]*Department of Chemistry, University of Colorado Boulder, CO 80309, USA*
[5]*Renewable and Sustainable Energy Institute, University of Colorado Boulder, CO 80309, USA*
[6]*EaStCHEM, School of Chemistry, Edinburgh University, David Brewster Road, Edinburgh EH9 3FJ, U.K.*
[7]*Molecular Foundry, Lawrence Berkeley National Laboratory, Berkeley, CA, USA.*
[†]*Present address: Stanford PULSE Institute and Linac Coherent Light Source, SLAC National Accelerator Laboratory and Stanford University, Menlo Park, CA 94025, USA.*
*Email: Quynh.L.Nguyen@colorado.edu

**Warm-dense matter (WDM) represents a highly-excited state that lies at the intersection of solids, plasmas, and liquids and that cannot be described by equilibrium theories. The transient nature of this state when created in a laboratory, as well as the difficulties in probing the strongly-coupled interactions between the electrons and the ions, make it challenging to develop a complete understanding of matter in this regime. In this work, by exciting isolated ~8 nm copper nanoparticles with a femtosecond laser below the ablation threshold, we create uniformly-excited WDM. Using photoelectron spectroscopy, we track the instantaneous electron temperature and directly extract the electron-ion coupling of the nanoparticle as it undergoes a solid-WDM phase transition. By comparing with state-of-the-art theories, we confirm that the superheated nanoparticles lie at the boundary between hot solids and plasmas, with associated strong electron-ion coupling. This is evidenced both by the fast energy loss of electrons to ions, and a strong modulation of the electron temperature induced by the variation in the nanoparticle volume. This work demonstrates a new route for experimental exploration of the exotic properties of WDM.**

Progress in several research areas depends on a detailed understanding of matter under extreme conditions of temperature and pressure. The "warm dense matter" (WDM) regime corresponds to matter with a density near those of solids and a temperature from ~10 K to ~10,000 K – a regime that cannot be described by equilibrium theories[1,2]. WDM lies at the heart of numerous unsolved problems in high-energy density physics[3,4], fusion energy sciences[5], planetary sciences and stellar astrophysics[6,7]. Enabled by advances in laser technology, the last decade has seen rapid progress in the ability to make and interrogate WDM in the laboratory[8-17]. However, despite breakthroughs in the past decade, it remains very challenging to accurately characterize the nature of interactions within WDM, particularly because of the transient nature of WDM creation in the lab, where dynamics can span from femtosecond timescales on up. This makes it difficult to validate advanced theories.

Previous studies used high-power lasers, pulsed power, and ion beams at mid and large-scale facilities to excite and probe WDM. These experiments have been limited to relatively low repetition rates (<120 Hz)[18] that influence what diagnostic measurements are possible. Additionally, the time resolution is often limited by the probing x-ray pulse duration or by streak camera diagnostics (>2 ps).[19,20] While ultrafast laser excitation at >1 kHz repetition rates can enable faster and more accurate measurements, the short penetration depth of laser light into materials, exponentially-decaying over ~10 nm, has represented a severe limitation to this approach. In a standard flat solid-target geometry, this means that the excitation and material phase of the WDM system changes dramatically with depth[1,2,21]. Moreover, the high laser energy necessary to compress or superheat a sample typically causes irreversible damage, requiring a fresh sample for every laser shot[9,21].

To create a more-uniformly heated sample, x-ray heating of thin films can be done, but requires a large-scale XFEL facility with limited repetition-rate and associated complexities.[10] Here, we present a novel approach for overcoming these long-standing challenges. Using intense ultrafast laser pulses, we uniformly heat a sample of isolated ~8 nm metallic nanoparticles to just below the ablation threshold. The nanoparticles are generated by a continuously operating magnetron sputtering source (see SM-S1.1), allowing for kHz repetition-rate data acquisition. We probe the resulting WDM dynamics using velocity-map-imaging photoelectron spectroscopy, which enables uniquely accurate measurements of this exotic state. We can directly measure the instantaneous electron temperature, and use this to accurately extract the electron-ion coupling ($G_{ei}$) dynamics and hot electron cooling in WD-nanomatter for the first time. We vary the excitation laser fluence to investigate the $G_{ei}$ dynamics below and above the melting threshold of the particles to characterize their properties in both solid and WDM phases, including the $G_{ei}$, decay rates, and phonon excitations. By comparing with theories of highly-excited strongly-interacting matter, we find that the state of matter we produce lies at the boundary between hot solids and plasmas. This is evidenced by the fast energy loss of electrons to ions, strong modulation of the electron temperature by variation in volume of the nanoparticle, and the existence of a threshold fluence associated with a change in electron-ion couplings. These results open a new avenue for experimental WDM research and inform current and future theories.

We use a magnetron sputtering source to produce ligand-free, isolated nanoparticles (NPs) in the gas phase (Fig. 1, see SM-S1.1). The measured temperature before laser excitation is ~1,000 K (extracted in a similar fashion to Shi et al.[22]); thus, the NPs start as a conventional hot nanoparticles. We characterized their size and shape using Aberration-Corrected Scanning Transmission Electron Microscopy (AC-STEM). The source yields fresh NPs that are uniform in size and shape[23] for every laser shot, circumventing any issues due to sample degradation by the excitation laser. Most importantly, the size of the NP (8.2 ± 1.1 nm, see SM-S1.2) is smaller than the absorption depth (~13 nm) at the excitation wavelength of 790 nm ( 1.5 eV photon energy), enabling homogenous heating and rapid thermalization of the electrons to a hot Fermi-Dirac distribution throughout the sample, and a near-complete elimination of effects due to spatial diffusion of heat at later times (Fig. 1a). The nanoparticle is first excited with an intense 790-nm pulse to excite electrons above the Fermi level, and then probed using a 395-nm pulse to ionize the atoms and capture the resulting electron (and phonon) dynamics (see SM-S2). We use laser intensities below the explosion threshold[24,25] to uniformly heat the nanoparticle. After ionization, the resulting photoelectrons are guided to a microchannel plate (MCP) with phosphor screen detector, and the images are captured by a camera.

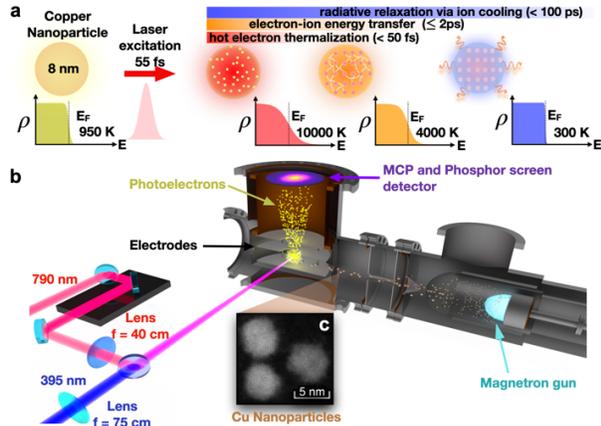

**Fig. 1 | Exciting and probing dynamics of warm dense CuNPs.** (**a**) After excitation by a 790-nm pulse, hot electrons thermalize on timescales <50 fs. This is followed by energy transfer from the electron bath to the lattice on timescales up to 2 ps[26-28]. The electron Fermi distribution (electron density ($\rho$) versus energy (E)) is shown, where $E_F$ is the Fermi energy. (**b**) Experimental apparatus for *in-vacuo* ultrafast photoelectron spectroscopy of CuNPs. A modified magnetron sputtering source produces monodisperse, highly pure CuNPs. Subsequently, they fly through a series of differential pumping stages and enter the interaction region where they interact with a pump-probe delay scheme. The photoelectrons are guided onto an MCP. (**c**) The lower inset shows a dark-field AC-STEM image of the produced CuNPs.

## Hot electron temperature measurement and modeling of the electron-ion coupling.

By illuminating the CuNPs with a moderately intense laser field ($10^{12}$ W/cm$^2$) below the ablation threshold, electrons are excited from the occupied to unoccupied states via inverse Bremsstrahlung absorption. Very fast electron-electron scattering gives rise to a hot thermalized Fermi-Dirac distribution (within ~50 fs)[26-28] with a very high temperature of ~10,000 K, that is initially much higher than that of the lattice (~1,000 K, see Fig. 1a). Next, scattering of hot electrons with ions becomes the dominant relaxation mechanism, which leads to excitation of phonon modes within ~200 fs[26-28], followed by non-thermal melting of the lattice, depending on the pump fluence. After a few picoseconds, the ions and the electrons reach thermal equilibrium at ~3000 K. As the particle transforms from solid-to-WDM, the lattice order vanishes and the electron charge density becomes more diffuse. Once the lattice exceeds the melting point of CuNPs ($T_{melt}$=1100K, see SM-S1.3) at the threshold (melting) fluence $F_{melt}$ = 107 mJ/cm$^2$ and higher, strong acoustic breathing modes are launched and modulate the volume of the CuNPs at ~3 ps and beyond (Fig. 2c). Radiative cooling of the superheated nanoparticle will then dominate after several hundred picoseconds.

First, to experimentally measure the instantaneous electron temperature (see Shi *et al.*[22]), we fit the photoelectron energy distribution curves (PEDC) to a hot Fermi-Dirac distribution at different time delays after excitation (Figs. 2a & b). Using the same procedure, we obtained the dynamical electron temperature distributions for different laser fluences from 80 – 147 mJ/cm$^2$ (Fig. 2c).

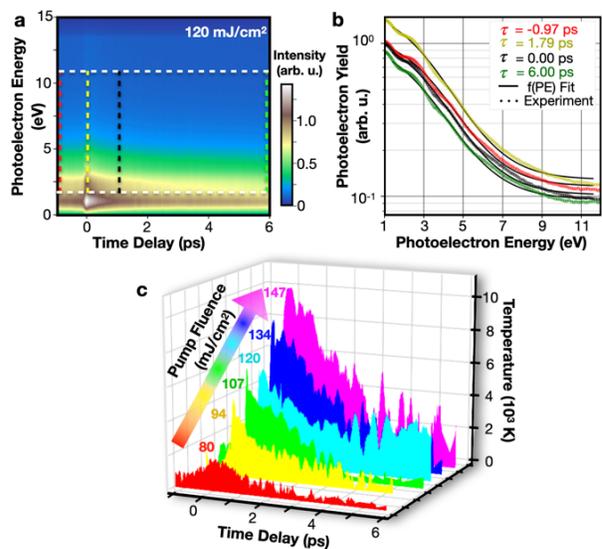

**Fig. 2 | Evolution of the hot electron temperature.** (**a**) Photoelectron energy as a function of pump-probe delay for pump laser fluence of 120 mJ/cm$^2$. Dashed colored lines indicate the PEDC at different time delays. The horizontal dashed white lines indicate the energy range for the fitting. (b) PEDC at time delays that correspond to the colored-dashed lines in panel (a) on a semilog-scale. The solid black lines show a numerical fit of the data to a thermal distribution, which is used to extract the electron temperature at each time delay. (c) Electron temperature profiles as a function of time delay for pump fluences that range from below to above $F_{melt}$ = 107 mJ/cm$^2$.

Next, we used the following modified-TTM to extract the electron-phonon coupling $G_{ei}$ (Eqs. 1 and 2)[29] at varying laser fluences ($F$) (see SM-S3-8 for further details). $G_{ei}$ controls the energy transfer rate from electrons to the phonons/ions.

$$C_e(T_e)\frac{\partial T_e}{\partial t} = \nabla(\gamma \nabla T_e) - G_{ei}(T_e - T_i) + \int dr\, \widetilde{p_e} \nabla \cdot v + sS_{laser}(t) \quad (1)$$

$$C_i \frac{\partial T_i}{\partial t} = \nabla(\gamma \nabla T_i) + G_{ei}(T_e - T_i) \quad (2)$$

where $C_e$ and $C_i$ are the electron and ion heat capacities (see SM-S5-7), $T_e$ and $T_i$ are the electron and ion temperatures, $\gamma$ is the thermal conductivity, $S_{laser}(t)$ is the laser heat source, $\widetilde{p_e}$ is the effective electron pressure, $\nabla \cdot v$ is the divergence of the electron velocity, and $s$ is the surface enhancement factor (see SM-S3). As a first-order approximation, we assume that $G_{ei}$ in Eq. 1 is constant over the time-delay range of our measurements and set the electron pressure term, $\int dr\, \widetilde{p_e} \nabla \cdot v$, to zero, which is denoted as simple-TTM. We fit the simple-TTM solution to the measured $T_e$ by varying $G_{ei}$ (see SM-S8, Fig. 3a) and repeat this procedure for all laser fluences. The extracted $G_{ei}$ are displayed as a function of peak $T_e$ (Fig. 3c).

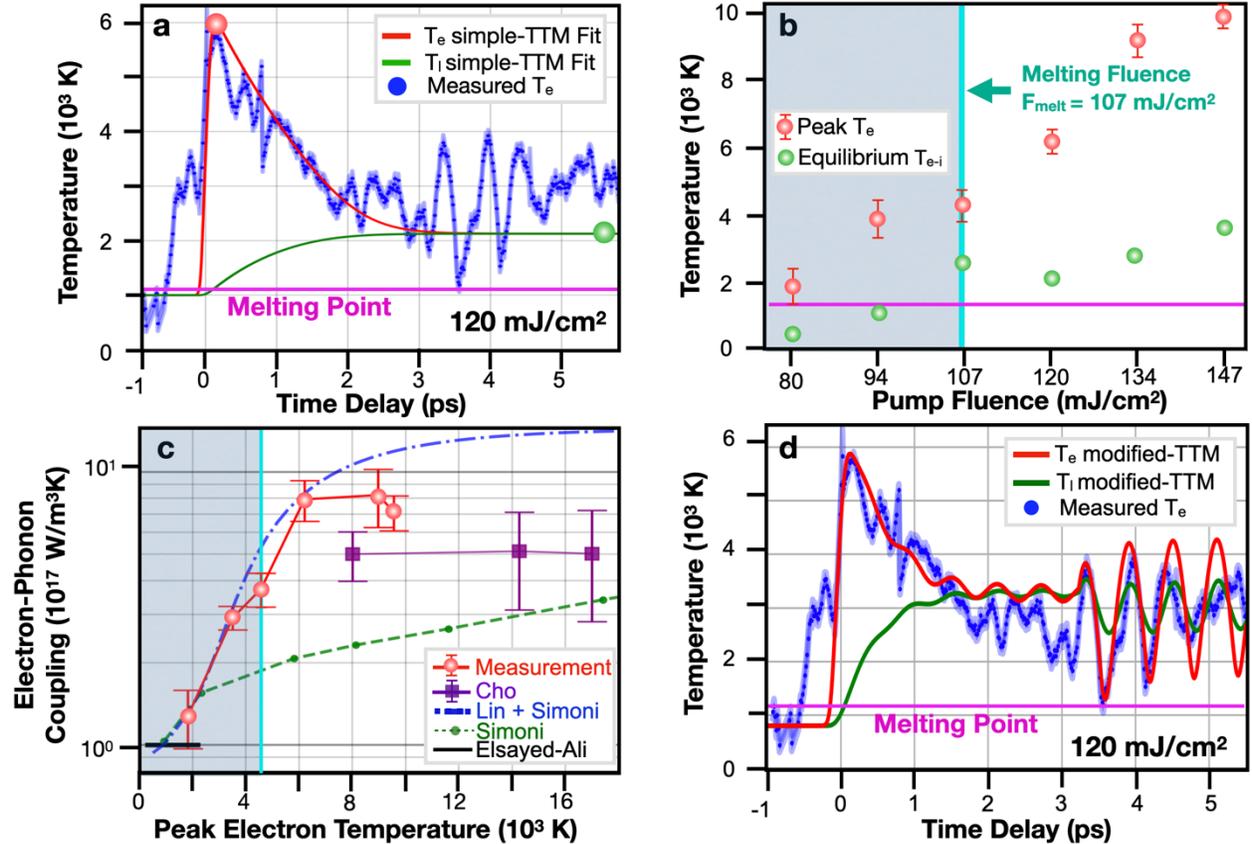

**Fig. 3 | Electron-ion (phonon) couplings in warm-dense CuNPs. (a)** Simple-TTM (without pressure) fit applied to the $T_e$ profile at 120 mJ/cm² to extract $G_{ei}$. $T_e$ (red) and $T_i$ (green) are the fits obtained for the electron and ion temperature profiles, where the error bars for $T_e$ are in shaded blue. The magenta solid line indicates the melting temperature for CuNPs, $T_{melt}$ = 1100 K. The peak-$T_e$ (red circle) and equilibrium electron-ion temperature $T_{e-i}$ (green circle) are determined from the $T_e$ and $T_i$ fits. **(b)** The extracted peak-$T_e$ and $T_{e-i}$ are shown for $F$ = 80 to 147 mJ/cm², where the lattice temperature exceeds $T_{melt}$ at $F_{melt}$ = 107 mJ/cm². The solid and WDM regimes are separated by a turquoise vertical line, where the blue region represents $T_i < T_{melt}$ and white region shows $T_i > T_{melt}$. We use this same notation in panel (c). **(c)** $G_{ei}$ fitted from the simple-TTM and displayed as a function of the peak-$T_e$ corresponding to each fluence. Theoretical $G_{ei}$ are calculated using Lin+Simoni (blue), and Simoni (green) models[34]. The measurements

for flat-CuWDM (purple)[31] and solid Cu[33] (black) are included for comparison. (d) The obtained $G_{ei}$ from simple-TTM and measured phonon mode (1.9 THz) are implemented into the modified-TTM for the same data shown in panel (a). $T_e$ (red) and $T_i$ (green) profiles are predicted using the semi-empirical modified-TTM. The error bars in all panels indicate 95% confidence interval.

As the solid-WDM phase transition occurs, the electrons become strongly degenerate, as more d-electrons are excited and start participating in the ion-electron energy exchange, leading to an increase in $G_{ei}$[30]. Interestingly, we observe a stronger variation for $G_{ei}$ than previously reported for ~70-nm thick Cu foil[20,31,32], which plateaus at ~4 to 6×10$^{17}$ W/Km$^3$ from 10,000 to 18,000 K while our measured $G_{ei}$ increases rapidly from 1.3 ± 0.3 ×10$^{17}$ to 8.2 ± 0.9 ×10$^{17}$ W/Km$^3$ between $T_e$ ~1,800 to 6,000 K, beyond which it plateaus (Fig. 3c). Our measured $G_{ei}$ is twice as high as those reported in prior works for flat-CuWDM, which used fluences that are two times higher. While these prior studies focus on the WDM regime, we vary the excitation laser fluence to observe variation in the properties of the CuNPs as they undergo a phase transformation from solid-to-WDM.

We compare our measurements with the theoretical models developed by Simoni et al.[34] for bulk WD-plasmas. We numerically evaluated $G_{ei}$ (see SM-S8) using this approach with the electronic and ionic structures calculated with Density Functional Theory based quantum molecular dynamic simulations[34] (QMD). Both theoretical and experimental $G_{ei}$ increase at higher $T_e$; however, the measured values increase more strongly with temperature (Fig. 3c). Meanwhile, we also computed the temperature dependent $G_{ei}$ using the model of Lin et al.[35,36]. In which, $G_{ei}$ is expressed as the product of the coupling ($G_{ei}^0$) at room temperature and a correction factor that depends on the density of states (DOS) to account for the number of thermally excited electrons. The Simoni approach accurately reproduced our measured $G_{ei}$ at $T_e$ ~1,800 K and the measured coupling for solid Cu[20,33] at room temperature, so-called $G_{ei}^0$. We set $G_{ei}^0$ in the Lin model to this value, the so-called "Lin+Simoni" model, and obtained excellent agreement with our measurements (Fig. 3c).

For the case of WD-nanoparticles, we note that additional relaxation mechanisms likely play a crucial role, that are not usually considered in existing models for flat-Cu[34-37]. These include excitation of volume-modulated phonon modes and quantum confinement effects for NPs. We observe a strong modulation that occurs after 3 ps at $F \geq F_{melt}$=107 mJ/cm$^2$ that signals a phase transition. This is due to changes in the particle volume as the lattice order vanishes[12,14] that cause the electrons to be more diffuse as the particle melts (see SM-S11-13). The simple-TTM model disagrees strongly with the measured $T_e$ for $F > F_{melt}$ and after 3 ps for $F > F_{melt}$, which indicates that the volume-pressure effects become significant as the particle transforms from solid-to-WDM. Furthermore, our QMD calculations predict a prominent pressure increase (>150 kBar) associated with this phase change (see SM-S12).

To understand the influence of the volume variation on $T_e$, we introduce a pressure-volume term that accounts for the energy change due to modulation in the volume of the nanoparticle, $\int dr\, \widetilde{p_e}\nabla \cdot v$ (Eqs. 1 & 2, see SM-S11 and Fig. 3c & d). We performed a Fourier Transform of the temperature profile and determined that the frequency of the large modulation is 1.9 THz, that is also in agreement with our MD simulations (see SM-S10). We note that similar oscillations have recently been observed in 2D materials at high laser fluences, where $T_e$ is modulated by coherent phonons associated with a charge density wave distortion[38]. Both represent evidence for very strong electron-phonon/electron-ion couplings. These surprising phenomena that we observed experimentally may be uniquely associated with the intrinsic properties of isolated CuNPs, where hot electrons are confined more effectively than in a flat-target geometry, and thus have not been previously reported.[12,20,31,32]

**Electron-ion coupling dynamics.**
We applied a phenomenological exponential model convoluted with the instrument response function (IRF) to further understand coupling to the ions as well as radiation loss (see SM-S14, S15). To directly compare the electron-ion energy transfer rate ($k_{ei}$) of CuNPs as the particle crosses the solid-WDM phase transition, we compare the measured $T_e$ profiles and cooling rates for fluences below (94 mJ/cm2) and above (120 mJ/cm$^2$) $F_{melt}$ at 107 mJ/cm$^2$ (Fig. 4a & b). The fitted 1/e time constants ($\tau_{ei} = 1/k_{ei}$) for the temperature profiles in the solid and WDM regimes are shown in Fig. 4c. The phase transformation leads to modifications in the electron-DOS and the heat capacity that influence $k_{ei}$. At higher $T_e$, the temperature dependence of $C_e$ reduces $k_{ei}$[39] while $G_{ei}$ exhibits an opposite effect. Within the solid and WDM regimes, $\tau_{ei}$ increases with higher $T_e$, which indicates that the $C_e$ effect dominates over the $G_{ei}$ influence (Fig. 4c). However, $C_e$ drops at the solid-WDM boundary, $F_{melt}$, and enhances $k_{ei}$. As a result, $\tau_{ei}$ exhibits a sudden drop, or a kink, at $F_{melt}$ that signals the phase transition. Our extracted $\tau_{ei}$ for the WD-nanomatter at $T_e$ ~10,000 K (1.5 $\pm$ 0.1 ps) is ~1.3 ps shorter than that reported for flat-CuWDM (2.8 $\pm$ 0.4 ps)[31]. This is likely due to the additional thermalization mechanisms occurring in the WD-nanomatter. For example, the pressure change connected to the time variation in the nanoparticle volume can speed up energy redistribution in the superheated CuNPs. This effect has also been theoretically predicted to accelerate the electron-ion heat exchange process in WD plasmas[40,41] and observed for semiconductors and semimetals.[42-44]

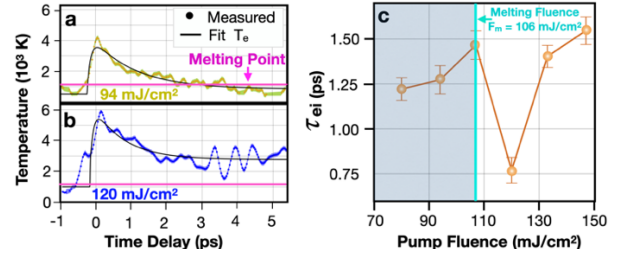

Fig. 4 | **Hot electron dynamics of CuNPs.** Extracted temporal evolution of $T_e$ at pump fluences of (a) 94 mJ/cm$^2$ (below $F_{melt}$) and (b) 120 mJ/cm$^2$ (above $F_{melt}$) to compare the volume-surface effects for CuNPs crossing the phase transition temperature. The solid black lines represent a fit to the data convoluted with the IRF of 99 $\pm$ 1 fs (see SM-S14), and are used to determine $\tau_{ei}$. The magenta solid horizontal line indicates $T_{melt}$ for CuNPs. (c) Extracted $\tau_{ei}$ for all pump fluences with error bars within 95% confidence intervals. The turquoise vertical line indicates $F_{melt}$ at 107 mJ/cm$^2$.

In conclusion, we have developed a unique method for probing the strong electron-ion coupling that is present in uniformly-excited WDM – in this case laser-excited CuNPs. By tracking the instantaneous hot electron temperature at varying pump fluences as a function of time, we can precisely extract the $G_{ei}$ as the CuNPs undergo a phase transformation. The measured $G_{ei}$ attains a maximal value of $(8.2 \pm 0.9) \times 10^{17}$ W/Km$^3$ over the transient electron temperature range from 1,800 to 10,000 K, which is higher than previously measured. Additionally, we experimentally and theoretically observe a large modulation on picosecond timescales in $T_e$ when the particle melts, as a result of the dramatic increase in the pressure accompanying a solid-WDM transformation. This study introduces a generalizable approach to the study of WDM phenomena across a wide range of different materials via the use of isolated nanoparticles. Finally, our direct measurements of electron energy loss can validate advanced theories that can include effects of particle-size, laser excitation parameters, pressure change, and strongly-coupled phonon modes on the electron-ion coupling and cooling pathways.

This work was supported by the DOE Office of Basic Energy Sciences Award No. DE-FG02-99ER14982. Q.L.D.N., J.L.E., N.J.B., and A.N.G. were also supported by the National Science Foundation Graduate Research Fellowship (DGE–1144083). S.Y. acknowledges support from the Facility for Electron Microscopy of Materials at the CU Boulder (CU FEMM). E.E.B.C acknowledges support of a JILA visiting fellowship. J.S. and J.D. were supported by the Laboratory Directed Research and Development program of Los Alamos National Laboratory under project number 20200074ER. This work was additionally supported by the user program of the Molecular Foundry, a DOE Office of Science User Facility supported by the Office of Science of the U.S. Department of Energy under Contract No. DE-AC02-05CH11231. This research used resources of the National Energy Research Scientific Computing Center, a DOE Office of Science User Facility supported by the Office of Science of the U.S. Department of Energy under Contract No. DE-AC02-05CH11231. We thank Joshua Knobloch and Christian Gentry for fruitful discussions.
**References:**

1. At the Frontier of Scientific Discovery, U.S. DOE Report of the Panel on Frontiers of Plasma Science, 2017.
2. Lee, R. W. *et al.* Finite temperature dense matter studies on next-generation light sources. J. Opt. Soc. Am. B **20**, 4 (2003)
3. Lebedev, S. *High Energy Density Laboratory Astrophysics* (Springer,2007).
4. Glenzer, S. H. *et al.* Symmetric inertial confinement fusion implosions at ultra-high laser energies. *Science* **327**, 1228–1231 (2010).
5. Nguyen, J. H. *et al.* Melting of iron at the physical conditions of the Earth's core. *Nature* **427**, 339 (2004).
6. Guillot T. "Interiors of giant planets inside and outside the solar system", *Science* **286**, 72 (1999).
7. Heinonen, R. A. *et al.* "Diffusion coefficients in the envelopes of white dwarfs", *The Astrophysical J.* **896**:2 (2020).
8. Ernstorfer, R. *et al.* The formation of warm dense matter: experimental evidence for electronic bond hardening in gold. *Science* **323**, 1033–1037 (2009).
9. Nagler, B. *et al.* Turning solid aluminium transparent by intense soft X-ray photoionization. *Nat. Phys.* **5**, 693-696 (2009).
10. Vinko, S. M. *et al.* Creation and diagnosis of a solid-density plasma with an X-ray free-electron laser. *Nature* **482**, 59-62 (2012).
11. Fäustline, R. F. *et al.* Observation of Ultrafast Nonequilibrium Collective Dynamics in Warm Dense Hydrogen. *Phy. Rev. Lett.* **104**, 125002 (2010).
12. Mahieu, B. *et al.* Probing warm dense matter using femtosecond X-ray absorption spectroscopy with a laser-produced betatron source. *Nat. Comm.* **9**, 3276 (2018).
13. Falk, K. "Experimental methods for warm dense matter research", *High Power Laser Science and Engineering* **6**, 59 (2018).
14. Fletcher, L. B. *et al.* "Ultrabright X-ray laser scattering for dynamic warm dense matter physics" *Nat. Photon.* **9**, 274–279 (2015).
15. Hayes, A. C. *et al.* "Plasma stopping-power measurements reveal transition from non-degenerate to degenerate plasmas" *Nat. Phys.* 16, 432-437 (2020).
16. Zylstra, A.B. *et al.* "Measurement of charged-particle stopping in warm dense matter", *Phys. Rev. Lett.* **114**, 215002 (2015).
17. F. Graziani *et al.* editors, *Frontiers and Challenges in Warm Dense Matter* (Springer-Verlag, Heidelberg, 2014).
18. Glenzer, S. H. *et al.* Matter under extreme conditions experiments at the Linac Coherent Light Source. *J. Phys. B: At. Mol. Opt. Phys.* 49, 092001 (2016)
19. Feng, J. *et al.* A grazing incidence x-ray streak camera for ultrafast, single-shot measurements. *Appl. Phys. Lett.* 96, 134102 (2010)
20. Cho, B. I. *et al.* Electronic Structure of Warm Dense Copper Studied by Ultrafast X-ray Absorption Spectroscopy. *Phys. Rev. Lett.* 106, 167601 (2011).
21. Dorchies, F. *et al.* Unraveling the Solid-Liquid-Vapor phase transition dynamics at the atomic level with ultrafast X-ray absorption near-edge spectroscopy. *Phys. Rev. Lett* 107, 245006 (2011)
22. Shi, X. *et al.* Ultrafast electron calorimetry uncovers a new long-lived metastable state in 1*T*-TaSe2 mediated by mode-selective electron-phonon coupling. *Sci. Adv.* 5, eaav4449 (2019).
7